\documentclass[11pt,a4paper]{article}
\usepackage{jstyle}
\usepackage{tikz-cd}
\usetikzlibrary {decorations.pathmorphing, decorations.pathreplacing, decorations.shapes}

\usepackage{parskip}
\hypersetup{
	colorlinks=true,
	allcolors=[rgb]{0.28, 0.24, 0.75}} 

\author[a,b,c]{Calvin Yi-Ren CHEN,}
\author[d]{\quad Euihun JOUNG,}
\author[a]{\quad Karapet MKRTCHYAN}
\author[d,e,f]{\quad and Junggi YOON}

\affiliation[a]{Theoretical Physics, Blackett Laboratory, Imperial College, London, SW7 2AZ, UK}
\affiliation[b]{Leung Center for Cosmology and Particle Astrophysics, Taipei 10617, Taiwan}
\affiliation[c]{Center for Theoretical Physics, National Taiwan University, Taipei 10617, Taiwan}
\affiliation[d]{
    Department of Physics,
    College of Science, 
    Kyung Hee University, Seoul 02447, Republic of Korea}
\affiliation[e]{Asia Pacific Center for Theoretical Physics, POSTECH, 77 Cheongam-ro, Nam-gu, Pohang-si, Gyeongsangbuk-do, 37673, Korea}

\affiliation[f]{School of Physics, Korea Institute for Advanced Study,
85 Hoegiro Dongdaemun-gu, Seoul 02455, Korea}

\emailAdd{cyrchen@ntu.edu.tw}
\emailAdd{k.mkrtchyan@imperial.ac.uk}
\emailAdd{euihun.joung@khu.ac.kr}
\emailAdd{junggi.yoon@khu.ac.kr}

\title{\centering
Higher-order chiral scalar \\
from boundary reduction of \\
3d higher-spin gravity
}

\abstract{ 
We use a recently proposed covariant procedure to reduce the Chern-Simons action of three-dimensional higher-spin gravity to the boundary, resulting in a Lorentz covariant action for higher-order chiral scalars. 
After gauge-fixing, we obtain a higher-derivative action generalizing the $s=1$ Floreanini-Jackiw and 
$s=2$ Alekseev-Shatashvili actions to arbitrary spin $s$.
For simplicity, we treat the case of general spin at the linearized level, while the full non-linear asymptotic boundary conditions are presented in component form for the $SL(3,\mathbb R)$ case.
Finally, we extend the spin-3 linearized analysis to a background with non-trivial higher-spin charge and show that it has a richer structure of zero modes.
}

\begin{document}

{\phantom{.}\vspace{-2.5cm}\\\flushright Imperial-TP-KM-2025-01\\ }

\maketitle

\section{Introduction}

Chiral degrees of freedom arise ubiquitously in theories of both fundamental interactions and emergent phenomena.
Instead of viewing them as troublesome special cases for which standard descriptions fail, one can take the alternative perspective that they are fundamental building blocks: They provide the simplest examples of field theories describing bosons with one-derivative free equations of motion similar to those of fermions. 
The methods developed for these degrees of freedom can then be used more generally.

One of the most natural descriptions of chiral modes is to present them as edge modes of topological theories in one higher dimension (see \textit{e.g.} \cite{Moore:1989yh,Witten:1996hc}). 
The typical and most-studied example of such a system is Chern-Simons (CS) theory on a three-dimensional manifold with Lorentzian boundary\footnote{Even though the bulk CS theory is topological, a metric structure can be introduced via the boundary term.}, where the chiral edge modes live. 
The latter can be described by a certain chiral Wess-Zumino-Witten\footnote{For early works on the (non-chiral) Wess-Zumino-Witten models, one can consult \textit{e.g.} \cite{Knizhnik:1984nr,Felder:1988sd}.} (WZW) model \cite{Sonnenschein:1988ug,Witten:1991mm}. 
Typically, these chiral modes are formulated without manifest Lorentz covariance \cite{Marcus:1982yu,Floreanini:1987as,Henneaux:1988gg} typically in favor of manifest T-duality (see \textit{e.g.} \cite{Tseytlin:1990nb,Tseytlin:1990va,Alvarez:1994dn,Klimcik:1995dy}). 
Several Lorentz covariant formulations are also available by now \cite{Pasti:1996vs,Sen:2015nph,Mkrtchyan:2019opf} --- see \cite{Evnin:2022kqn} for a review. 
Remarkably, a prescription for generating boundary Lorentz covariant actions from bulk topological field theories was recently shown in \cite{Arvanitakis:2022bnr}. 
The resulting covariant actions are of the type introduced in \cite{Mkrtchyan:2019opf}. 
This previously allowed some of the present authors to find a natural generalization of the method in \cite{Arvanitakis:2022bnr} to arbitrary dimensions and arbitrary rank (Abelian) forms, including their Abelian interactions \cite{Evnin:2023ypu}, reproducing the results of \cite{Avetisyan:2022zza,Mkrtchyan:2022xrm}. 
Note that this reduction is tied to a given background and therefore insensitive to effects from non-trivial topologies in CS (or BF) theories --- see \textit{e.g.} \cite{Porrati:2021sdc} for a discussion on such effects. 
Ultimately, the reduction leads to a boundary classical field theory described by self-duality equations and their non-linear generalizations, which shall be the interest of this work.

Chern-Simons actions with non-compact gauge groups are known to describe theories of (higher-spin) gravity \cite{Achucarro:1986uwr,Witten:1988hc,Blencowe:1988gj,Vasiliev:1989qh} in three dimensions, including their color decorations \cite{Gwak:2015vfb,Gwak:2015jdo,Gomis:2021irw}. 
Similarly to their counterparts with compact gauge groups, the dynamical degrees of freedom of these theories also live on the boundary. 
The dynamics of these boundary degrees of freedom can also be described by a boundary action.
Precisely by using the technique introduced in \cite{Arvanitakis:2022bnr} supplemented with boundary conditions necessary due to the gravitational nature of these theories, we will derive these boundary actions. 
In other words, this is effectively an extension of a covariant action for the chiral WZW model given in \cite{Arvanitakis:2022bnr} to the gauged chiral 
WZW model (with non-compact symmetry) \cite{Gawedzki:1988hq,Gawedzki:1988nj}.

For pure gravity (with $SL(2,\mathbb R)$ gauge group), the reduced action we end up with is equivalent to the Lorentz covariant counterpart of two copies of the Alekseev-Shatashvili (AS) action \cite{Alekseev:1988ce,Alekseev:1988tj,Alekseev:1988vx} --- the equivalence of 3d CS gravity and AS theory was previously shown in \cite{Cotler:2018zff}.
See also \cite{Afshar:2024ess, Detournay:2024gth}
for the $SL(2,\mathbb R)\times U(1)$
extension.
We generalize this reduction procedure to higher-spin (HS) fields and obtain a covariant action for the ``HS generalization'' of AS theory.
At the linearized level, we perform this around an arbitrary gravitational saddle, in order to find a manifestly Lorentz covariant action for higher-order chiral scalars.
After gauge fixing these theories (at the price of breaking the manifest Lorentz covariance) we obtain a Floreanini-Jackiw (FJ) \cite{Floreanini:1987as} type action for the higher-order chiral scalars, describing the $SL(2,\mathbb R)$ representation with lowest weight $\D=s$.
Its kinetic term admits an interesting factorization property, from which we can deduce the existence of additional zero modes at which there is a gauge symmetry enhancement. 
In the nonlinear case, the manifest Lorentz covariant action can be written for any gauge group such as $SL(N,\mathbb R)$ since the asymptotic AdS conditions can be implemented as constraints --- this results in the covariant version of the gauged chiral WZW action.
Complications arise when we fix a manifestly Lorentz-breaking gauge and solve the constraints.
We revisit the simplest example of $SL(3,\mathbb R)$ CS actions, which have been considered in earlier work \cite{Bershadsky:1989mf,Marshakov:1989ca,Bilal:1990wn} (see also \cite{Li:2015osa} for a more recent account and \cite{Merbis:2023uax} for a similar reduction procedure to the current work) in order to highlight that the resulting boundary theory can be viewed as an interacting theory of a higher-order chiral scalar with the lowest-weight $\D=3$.
In principle, the same can be done for the $SL(N, \mathbb R)$ case. 
The resulting theory is closely related to Toda theory (see {\it e.g.} \cite{Gonzalez:2014tba,Ma:2019gxy,Lomartire2023} for the recent discussions) or $\cW_N$-gravity \cite{Hull:1993kf} similar to how AS theory is related to Liouville theory \cite{Coussaert:1995zp}.

Another benefit of our discussion in comparison to earlier literature is a more natural separation of fluctuation modes from the background: not only do we work with the most generic (asymptotically AdS) gravitational background parameterized by $\cL$ \cite{Brown:1986nw,Banados:1998gg} with a particular
care on its zero mode like
in \cite{Merbis:2023uax}, we also assess some implications of the spin-3 charge $\cW$.
The more interesting example of HS black holes is left for future investigation, as it requires us to modify crucial elements of our current work, namely the boundary condition (the asymptotic AdS condition) and the boundary term.
We expect there to be a different boundary term which is consistent with HS backgrounds and able to reproduce the same results for the Brown-Henneaux type boundary conditions.
We hope to revisit this possibility in future.

Organization of the current paper is as follows.
We begin with the boundary reduction of the linearized theory, namely the 3d Fronsdal action, in Section \ref{sec: fronsdal}.
After briefly reviewing the CS description of (HS) gravity, we show that the bulk equations admit nontrivial solutions determined by their boundary value, namely the edges modes.
We then move to the topological derivation as laid out in \cite{Arvanitakis:2016vnp} of the Lorentz covariant boundary action and supplement it with the asymptotic AdS conditions to obtain an action for higher-order chiral scalars.
Upon gauge fixing, we see that the kinetic term of this action can be written in factorized form, thus exhibiting points with gauge symmetry enhancement.
In Section \ref{sec: nonlinear}, we extend the analysis to the nonlinear theory with $SL(N,\mathbb R)$ gauge group.
The resulting Lorentz covariant boundary action is the Lorentz covariant counterpart of the gauged chiral WZW model.
For the pure gravity case with $N=2$, we fix a manifest-Lorentz covariance breaking gauge and reduce the action to the AS action.
For HS gravity with $N=3$, the reduction provides a set of nonlinear equations as the asymptotic AdS conditions, demonstrating that the result is a nonlinear action for the higher-order chiral scalar.
Finally, in Section \ref{sec: HS charge}, we consider a background with nontrivial HS charge $\cW$ together with the gravitational background $\cL$, which can be viewed as a generalized conical defect.
Again focusing on the linearized fields, we identify the kinetic operator and show a richer structure of the gauge symmetry enhancement points.

\subsection*{Conventions}

We work in units where $\hbar, c =1$ and in mostly-plus signature $(-,+,\dots,+)$.
It will sometimes be convenient to use null coordinates, which we will take to be $x^{\pm} = x^{1} \pm x^{0}$.
We define the AdS length scale $\ell$ in terms of the cosmological constant $\Lambda$ via $\Lambda = -1/\ell^{2}$.
In the context of the Chern-Simons description of (HS) gravity in $\mathrm{AdS}_{3}$, Newton's constant $G$ is related to the Chern-Simons level $k$ by $G = \ell/4k$.
In dealing with HS fields, we will also denote traceless symmetrization of indices using curly brackets.

\section{Edge mode action of Fronsdal fields}
\label{sec: fronsdal}

In this section, we will derive covariant actions for the edge modes of HS fields in $\mathrm{AdS}_{3}$.
We start with a brief review of the CS formulation of Fronsdal fields and discuss the equations of motion for the degrees of freedom.
We then perform the topological reduction of the action to the boundary, to obtain a covariant action for the boundary degrees of freedom.
This is compared to the actions in the form of 
Pasti-Sorokin-Tonin (PST) \cite{Pasti:1996vs} and FJ \cite{Floreanini:1987as}.

\subsection{Bulk action}

Let us briefly review the CS formulations of spin-2 and HS fields in the bulk of $\mathrm{AdS}_{3}$.

\subsubsection{Spin-2}

Let us first consider the Einstein-Hilbert action with negative cosmological constant $\Lambda = -1/\ell^{2}$ defined on a manifold $M$ with timelike boundary $\partial M$. 
Up to boundary terms, this action can be recast in terms of two copies of the CS action with Lie algebra $\mathfrak{so}(1,2)$\,: 
\ba
    S_{\rm EH}[e,\o] \eq \frac{1}{8\pi G} \int_{M} e_a\wedge \left[\dd \o^{a}+\frac12\,\e^{a}{}_{bc}\left(\o^{b}\wedge \o^c + \frac{1}{3\ell^{2}}e^{b} \wedge e^{c}\right)\right] \nn
    \eq S_{\rm CS}[A]-S_{\rm CS}[\tilde A]\,,
\ea
where the local Lorentz $\mathfrak{so}(1,2)$ indices $a,b,c,\dots$ take value in $\{0,1,2\}$ with metric $\eta={\rm diag}(-1,+1,+1)$, and the Levi-Civita symbol is defined with $\epsilon_{012}=+1$.
The CS action is given by
\be
    S_{\rm CS}[A] = \frac{k}{4\pi} \int_M\tr \left(A\wedge \dd A+\frac23\,A\wedge A\wedge A\right),
    \label{eq: cs action}
\ee
where the CS level $k$ is related to Newton's constant by $k=\ell/4G$, and each of the $\mathfrak{so}(1,2)$ CS gauge fields, $A=A^a\,J_a$ and $\tilde A=\tilde A^a\, J_a$, are related to the spin connection and dreibein by
\be
	A^a=\o^a + \frac{e^a}{\ell}\,,\qquad  \tilde{A}^a=\o^a - \frac{e^a}{\ell}\,.
\ee
For the generators of $\mathfrak{so}(1,2)$, we will use the convention,
\begin{equation}
    \tr(J_{a}\,J_{b})=\frac12\,\eta_{ab}\,,\qquad [J_{a},J_{b}]=\epsilon_{ab}{}^{c}\,J_c \,.
\end{equation}

Up to boundary terms, the variation of the CS action, $\delta S_{\text{CS}} \propto	\int_M\tr\left[\delta A\wedge F\right]$,
gives the flat curvature equation $F=\dd A+A\wedge A=0\,$.
The flatness condition for the two $\mathfrak{so}(1,2)$ gauge fields are equivalent to the Cartan structure equations,
\be 
    \dd e^a+\e^a{}_{bc}\, \o^b\wedge e^c=0\,, \qquad \dd \o^a+ \frac12\,\e^{a}{}_{bc}
    \left(\o^b\wedge \o^c+
    \frac1{\ell^2}\,e^b\wedge e^c\right)=0\,.
    \label{ew eq}
\ee

Let us now consider fluctuations of $A^a$ around a background solution $\underline{A}^a$.
In terms of the gauge field, we consider
\be
    A^a =\underline{A}^a+a^a\,,\qquad \tilde{A}^a= \underline{\tilde{A}}^a+\tilde{a}^a\, 
\ee
so that the CS action can be expanded as 
\be
    S_{\rm CS}[\underline{A} + a]=S_{\rm CS}[\underline{A}] +\frac{k}{4\pi}\int_M \frac12\,a_a\wedge\Dd a^a+\frac{1}{6}\,\e_{abc}\,a^a\wedge a^b\wedge a^c\,,
    \label{eq: cs action expanded linear}
\ee 
where $\Dd a^a$ is the covariant derivative given by
\be
    \Dd a^a=\dd a^a+\e^{a}{}_{bc}\, \underline{A}^b\wedge a^c\,.
\ee
The expansion of $\tilde A^a$ follows similarly. 
We will further focus on the quadratic action and neglect the third-order terms. 
Since the covariant derivative $\Dd$ is nilpotent, the quadratic action has the same algebraic structure as an Abelian CS action.

\subsubsection{Higher-Spin}

Now we turn to HS gauge fields, namely the Fronsdal fields --- our conventions will mostly follow \cite{Campoleoni:2024ced}.  
In the frame-formulation for spin-$s$ Fronsdal fields in an Einstein background, the action reads
\begin{equation}
    \begin{aligned}
    & S_{\rm Fronsdal} =  \frac{1}{8\pi G}\int_M
    \bigg\{\varphi_{a_1\cdots a_{s-1}} \wedge \left[ \dd\omega^{a_{1} \cdots a_{s-1}} +(s-1)\epsilon^{bc (a_{1}} \underline{\omega}_{b} \wedge \omega^{a_{2}\cdots a_{s-1})}{}_{c}\right] \\
    & +\frac{s-1}{2}\epsilon_{bcd}\, \underline{e}^{b} \wedge \omega^{c}{}_{a_{2}\cdots a_{s-1}} \wedge \omega^{d a_{2} \cdots a_{s-1}} +\frac{s-1}{2\ell^{2}}\epsilon_{bcd}\, \underline{e}^{b} \wedge \varphi^{c}{}_{a_{2}\cdots a_{s-1}} \wedge \varphi^{d a_{2} \cdots a_{s-1}}
    \bigg\}
    \label{Fronsdal}
    \end{aligned}
\end{equation} 
where the one-form fields
$\varphi_{a_1\cdots a_{s-1}}$
and $\o_{a_1\cdots a_{s-1}}$
are traceless in fiber indices, and
$(a_1\cdots a_{s-1})$ is the symmetrization with total weight one. The $\underline{e}^a$ and $\underline{\o}^a$ satisfy the on-shell conditions \eqref{ew eq}.
Decomposing the two one-forms, namely the Fronsdal field $\varphi^{a_1\cdots a_{s-1}}$ and the (first) connection field $w^{a_1\cdots a_{s-1}}$, as 
\be
    a^{a_1\cdots a_{s-1}}=
    w^{a_1\cdots a_{s-1}}+\frac{1}{\ell}\varphi^{a_1\cdots a_{s-1}}\,,
    \qquad 
    \tilde a^{a_1\cdots a_{s-1}}=
    w^{a_1\cdots a_{s-1}}-\frac{1}{\ell}\varphi^{a_1\cdots a_{s-1}}\,,
\ee
the Fronsdal action \eqref{Fronsdal} can also be split, up to boundary terms, into two copies of Abelian CS actions:
\be 
    S_{\rm Fronsdal}[\varphi, w]=S_{\rm CS}[a]-S_{\rm CS}[\tilde a]\,.
\ee 
Here, the CS action is given by
\be 
    S_{\rm CS}[a^{a_1\cdots a_{s-1}}]= \frac{k}{4\pi}\int_M \frac12\,a_{a_1\cdots a_{s-1}}
    \wedge \Dd a^{a_1\cdots a_{s-1}}\,,
\ee 
with $\Dd a^{a_1\cdots a_{s-1}}
    =\dd a^{a_1\cdots a_{s-1}}
    +(s-1)\,\epsilon^{b}{}_c{}^{(a_1}\,
    \underline{A}_b\,a^{a_2\cdots a_{s-1})c}\,$ and we once again identify $k = \ell/4G$.
Further, by defining HS algebra generators $J_{a_1\cdots a_{s-1}}$ with
\be
    [J_a, J_{b_1\cdots b_{s-1}}]=
   (s-1)\,\e_{a(b_1}{}^c\,J_{b_2\cdots b_{s-1})c}\,,
    \qquad 
    \tr(J_{a_1\cdots a_{s-1}}\,J^{b_1\cdots b_{s-1}})
    =\frac1{2}\,\delta^{\{b_1}_{\{a_1}\cdots
    \delta^{b_{s-1}\}}_{a_{s-1}\}}\,,
    \label{Js gen}
\ee
with $\{a_1\cdots a_{s-1}\}$ denoting traceless symmetrization, the Abelian CS action can be expressed as
\be 
    S_{\rm CS}[a]=\frac{k}{4\pi}\int_M \tr\left(a
    \wedge \Dd a\right),
\ee 
with $a=a^{a_1\cdots a_{s-1}}\,J_{a_1\cdots a_{s-1}}$ and $\Dd a=\dd a+\underline{A}\wedge a + a\wedge \underline{A}$\,.
This describes a massless spin-$s$ field around an on-shell gravitational background, including linearized gravity
in the case of $s=2$.

\subsection{Equations of motion}

So far, we have been neglecting boundary contributions.
However, for a well-defined variational principle, we need to supply the action with an appropriate boundary term. 
We choose boundary terms with opposite signs for the two copies of the CS action:
\ba
    &&S_{\rm L}[a]=\frac{k}{4\pi}\left(\int_M \tr(a\wedge \Dd a)
    -\frac12\int_{\partial M}\tr(a\wedge \star a)\right)\,,
    \label{SL}
    \nn 
    &&  S_{\rm R}[\tilde a]= \frac{k}{4\pi}
    \left(\int_M \tr(\tilde a\wedge \Dd \tilde a)+\frac12\int_{\partial M}\tr(\tilde a\wedge \star \tilde a)\right)\,,
\ea
so that the total action becomes 
\be 
    S[a,\tilde a]=S_{\rm L}[a]-S_{\rm R} [\tilde a]\,.
\ee 
We will focus on $S_{\rm L}$ for now, since $S_{\rm R}$ can be treated in an analogous manner.

The variation of $S_{\rm L}$ is 
\be
   \frac{4\pi}{k}\,\delta S_{\rm L}= 2\int_M\tr(\delta a\wedge\Dd a) + \int_{\partial M}\tr[\delta a\wedge(a - \star a)]\,,
\ee 
which results in the boundary equation of motion,
\be
    (a - \star a)|_{\partial M}=0\,,
    \label{bd eq}
\ee
which can be equally interpreted as a Neumann boundary condition.

Note that we have introduced two reference metrics on $\partial M$: one through the covariant derivative $\Dd$ and the other through the Hodge dual. 
For the Hodge dual, we consider the boundary metric
given by the zweibein $\mathsf e^\a$,
\be
    \dd s^2=\eta_{\a\b}\,\mathsf e^\a\,\mathsf e^\b\,,
\ee 
with $\eta_{\oplus\ominus}=\frac12$ and $\eta^{\oplus\ominus}=2$.
The boundary Levi-Civita symbol $\e_{\a\b}$ is given by $\epsilon_{\oplus\ominus} = +\frac12$.
With these choices, the Hodge dual acts on the zweibein as
\be
    \star\, \mathsf e^\a=\e_\b{}^{\a}\,\mathsf e^\b\,, \qquad \star\,\mathsf e^{\oplus/\ominus}=\pm\,\mathsf e^{\oplus/\ominus}\,.
 \label{Hodge}
\ee

Let us first analyze the equations of motion to identify the content of the theory.
The equation of motion in the bulk $M$, $\Dd a=0\,,$ can be solved as $a=\Dd \varphi$ by nilpotency of $\Dd$.
Using the explicit Hodge dual map \eqref{Hodge}, the boundary equation of motion \eqref{bd eq} can be expressed  as
\be
     \Dd_{\ominus}\varphi\,|_{\partial M}=0\,,
    \label{chiral}
\ee
where we use the notation $\Omega_{\a}=\mathsf e_{\a}{}^\m \,\Omega_\m$ for any one form $\Omega$.

For the flat zweibein $\mathsf e^{\oplus/\ominus}\propto\dd x^{+/-}$ with $x^\pm=x^1\pm x^0$, the condition \eqref{chiral} reduces to the chirality condition $\Dd_-\varphi\,|_{\partial M}=0$.
Therefore, we find chiral field as the on-shell content of the theory. 
In the case of a genuine Abelian gauge field, this is the end of the story.
However, for Fronsdal fields, we should remember that $\varphi$ is a tensor field and hence contains many chiral boundary degrees of freedom. 
These degrees of freedom should be further reduced by an asymptotic AdS condition, which we shall consider later.

\subsection{Boundary reduction}

Setting aside the asymptotic AdS condition for now, we will now apply the reduction procedure introduced in \cite{Arvanitakis:2022bnr} to the Fronsdal action \eqref{SL} to obtain a manifestly Lorentz covariant action.
This covariant action provides the chirality condition \eqref{chiral} as its on-shell equation.

We begin with the decomposition of 
the one-form $a$, 
\be
    a=b+v\,\psi\,,
    \label{cov decomp}
\ee 
by employing a nowhere-null one-form $v$ with $\Dd v=0$\,. 
Note that $b$ and $\psi$ take value in the same tensor space as $a$,
and $\psi$ plays the role of a Lagrange multiplier imposing the constraint
\be
    v\wedge\Dd b=0\,.
\ee
This is satisfied when the one-form $a$ takes the form
\be
    a=\Dd\varphi+v\,\rho\,,
    \label{constr sol}
\ee
for a tensor field  $\rho$.
Note that the split of $a$ into $\Dd\varphi$ and $v\,\rho$ is not unique as it involves a redundancy, parameterized by gauge symmetry subject to the relation
\be 
    \Dd\delta \varphi+v\,\delta \rho=0\,.
\ee 
The variation $\delta\rho$ is completely determined by $\delta\varphi$ which satisfies
\be 
    v\wedge \Dd\delta\varphi=0\,.\label{gauge}
\ee 
By implementing \eqref{constr sol} in the action, we find
\ba
    S_{\rm L}\eq \frac{k}{4 \pi} \int_{\partial M} \tr\left[-\frac12\, (\Dd\varphi+v\,\rho)\wedge \star (\Dd\varphi+v\,\rho) - \Dd\varphi\wedge v\,\rho\right]\nn
    \eq -\frac{k}{8 \pi}\int_{\partial M}\tr\left[\Dd\varphi\wedge \star \Dd\varphi +2(\star v+v)\wedge \rho\,\Dd\varphi +\,\rho^2\,v\wedge \star v\,\right].
    \label{cov action}
\ea
In components, the above action reads
\be
	S_{\rm L} = - \frac{k}{2 \pi} \int_{\partial M}\dd^2 x\,\det\mathsf e\,
 \tr\left[ \Dd_\oplus\varphi\,\Dd_\ominus\varphi +2\,v_\oplus\rho\,\Dd_\ominus\varphi + \rho^2\,v_\oplus \, v_\ominus \right].
 \label{WZWM}
\ee
By integrating out $\rho$, we find the PST-type action,
\be 
    S_{\rm L} = - \frac{k}{4 \pi} \int_{\partial M}\dd^2 x\,\frac{\det \mathsf e}{v_\ominus}\,\tr\left[
    \e^{\a\b}\,v_\a\,\Dd_\b\varphi\,\Dd_\ominus\varphi\right].
     \label{PST type}
\ee
Note that for a flat metric, if we fix the gauge $v\propto \dd x^0$, we find the Floreanini-Jackiw (FJ) type action
\be
    S_{\rm L}= - \frac{k}{2 \pi} \int_{\partial M}\dd^2 x\, \tr(\Dd_1\varphi\,\Dd_- \varphi)\,.
    \label{FJ type}
\ee 
For an arbitrary zweibein $\mathsf e^\a$, the equation of motion of the action
\eqref{PST type} is
\be
\{\e^{\a\b}\,v_\a\,\Dd_\b,\Dd_\ominus\}\varphi=0\,.
\ee 
Since $[\e^{\a\b}\,v_\a\,\Dd_\b,\Dd_\ominus]=0$, the general solution is given by
\be
    \varphi=\varphi_{\rm ph}+\varphi_{\rm g}\,,
\ee 
where $\varphi_{\rm ph}$ and $\varphi_{\rm g}$ satisfy
\be
    \Dd_\ominus\varphi_{\rm ph}=0\,,\qquad 
    \e^{\a\b}\,v_\a\,\Dd_\b\,\varphi_{\rm g}=0\,.
\ee 
Note that $\varphi_{\rm g}$ can be precisely removed by gauge symmetry \eqref{gauge}. 
We therefore only have $\varphi_{\rm ph}$ as a dynamical field, whose equation of motion reduces to
\be 
    \Dd_\ominus \varphi=0\,.
\ee 
This is what we obtained directly from the CS action even before reduction to the boundary. 

\subsection{Asymptotic AdS Conditions}

We will now consider the asymptotic AdS condition.

In \cite{Banados:1998gg}, it was shown that the metric of an asymptotically AdS spacetime (in the sense of \cite{Brown:1986nw}) is generally given by 
\be
    ds^2
    =\ell^{2}\frac{(\dd x^++z^2\,\tilde\cL(x^-)\,\dd x^-)(\dd x^-+z^2\,\cL(x^+)\,\dd x^+)+\dd z^2}{z^2}\,
     \label{bg metric}
\ee
in lightcone Poincar\'e coordinates $x^\pm=x^{1}\pm x^{0}$ and $z=x^2$.
The corresponding frame fields and the spin connection are
\be
    \underline{e}^z=\ell\frac{\dd z}{z}\,, \qquad \underline{e}^+=\ell\frac{\dd x^++z^2\,\tilde \cL(x^-)\,\dd x^-}{z}\,,
    \qquad 
   \underline{e}^-=\ell\frac{\dd x^-
    +z^2\,\cL(x^+)\,\dd x^+}{z}
\ee 
\be 
    \underline{\o}^z=0\,, \qquad \underline{\o}^+=\frac{\dd x^+-z^2\,\tilde\cL(x^-)\,\dd x^-}{z}\,,
    \qquad 
    \underline{\o}^-=-\frac{\dd x^--z^2\,\cL(x^+)\,\dd x^+}{z}\,.
\ee
Combining the above, we find
the background solution
\ba
    \underline{A}^z \eq \frac{\dd z}{z}\,, \qquad \underline{A}^+=2 \frac{\dd x^+}{z}\,,\qquad \underline{A}^-=2\,z\,\cL(x^+)\,\dd x^+\,, \nn
    \underline{\tilde{A}}^z \eq -\frac{\dd z}{z}\,, \qquad \underline{\tilde{A}}^+=-2\,z\,\tilde\cL(x^-)\,\dd x^-\,,\qquad \underline{\tilde{A}}^-=-2\,\frac{\dd x^-}{z}\,.
    \label{AdS A}
\ea 
We take the zweibein $\mathsf e^\a$ associated with the boundary Hodge dual as the pullback of the 3d background metric \eqref{bg metric} to the boundary.
As we shall consider the case where the boundary is located at a constant $z$ surface with $z\to 0$, we take this zweibein to be flat.

For further analysis, it is useful to use the basis,
\be
    L_0=-J_2\,,
    \qquad
    L_{\pm1}=\pm2\,J_{\pm}=J_0\pm J_1\,,
\ee 
such that
\be 
    [L_m,L_n]=(m-n)\,L_{m+n}\,.
\ee
Then, we can expand the spin-2 field as
\be
    A = A^{\sst (+1)}\,L_{+1}
+A^{\sst (0)}\,L_0+A^{\sst (-1)}\,L_{-1}\,,
\ee
with 
\be 
    A^{\sst (0)}=-A^z\,,
    \qquad A^{\sst (\pm1)}
    =\pm\frac12\,A^{\pm}\,.
\ee 
The Fronsdal fields can also be expanded as
\be 
    a=\sum_{n=1-s}^{s-1} a^{\sst (s,n)}\,W^{(s)}_n\,,
\ee 
where the higher spin generators $W^{(s)}_n$ are 
\be 
    W^{(s)}_n\propto J_{\underbrace{+\cdots +}_{s-1+n}\underbrace{-\cdots -}_{s-1-n}}\,,
\ee 
and satisfy
\begin{subequations}    
    \begin{align}
        &[L_{m}, W^{(s)}_n]=(m (s-1)- n)\,W^{(s)}_{n\pm 1}\,,\\
        &\tr(W^{(s)}_n\,W^{(s)}_m) =\t(s)\,(-1)^m\,(s-1+m)!\,(s-1-m)!\,\delta_{m+n}\,,
    \end{align}
\end{subequations}
where $J_{\a_1\cdots \a_{2(s-1)}}$ is the $SL(2,\mathbb R)$ 
counterpart of 
the $SO(1,2)$-tensor $J_{a_1\cdots a_{s-1}}$ 
subject to the conditions 
\eqref{Js gen}.
So far $\t(s)$ is unconstrained, but it may be fixed by the choice of the HS algebra, after fixing the commutators of HS generators.

In the basis of $L_n$ and $W^{(s)}_n$, the bulk solution reads
\be 
    a^{\sst (s,n)}=\Dd \varphi^{\sst (s, n)}
    =z^{-n}\,\Dd\phi^{\sst(s, n)}\,,
\ee 
where $\varphi^{\sst (s, n)}(z,x)=z^{-n}\,\phi^{\sst (s,n)}(z,x)$
and $\Dd\phi^{\sst (n)}$ is given as
\be 
    \Dd\phi^{\sst (s,n)}=\dd \phi^{\sst (s,n)}
    +\left((s-n)\,\phi^{\sst (s,n-1)}
    +(s+n)\,\cL(x^+)
    \,\phi^{\sst (s,n+1)}
    \right)\dd x^+\,.\label{cov der}
\ee
For $v=\dd x^0$, the asymptotic AdS conditions for HS fields 
\cite{Henneaux:2010xg, Campoleoni:2010zq} (see also
\cite{Gaberdiel:2011wb, Campoleoni:2011hg}),
generalizing the gravity case \cite{Coussaert:1995zp},
are 
\be
    a^{\sst (s,n)}_1=\cO(z^{1-n}) 
    \qquad [2-s\le n\le s-1]\,,
    \label{asymp AdS}
\ee
and they are solved by $\phi^{\sst (s,n)}=\cO(z^0)$ obeying 
\be 
    \Dd_1\phi^{\sst (s,n)}(0,x)
    =0
    \qquad [2-s\le n\le s-1]\,.
    \label{const}
\ee 
This allows us to solve each
$\phi^{\sst (s,n)}$ in terms of $\phi^{\sst (s,n+1)}$
and $\phi^{\sst (s,n+2)}$ as
\be 
    \phi^{\sst (s,n)}
    =-\frac{\partial_1\phi^{\sst (s,n+1)}
    +(s+n+1)\,\cL\,\phi^{\sst (s,n+2)}}{s-n-1}\,,\label{recur}
\ee 
and hence 
all $\phi^{\sst (s,n)}$ with $n\ge 1-s$ can be expressed in terms
of $\phi^{\sst (s,s-1)}\equiv \phi^{\sst (s)}$.
Applying this result to the FJ-type action \eqref{FJ type} for a spin-$s$ field, we obtain 
\ba
    S_{\rm L}\eq 
 2\,\t(s)\,(2s-2)!(-1)^{s}\int_{\partial M}\dd^2 x\,
	\Dd_1\phi^{\sst (s,1-s)}\,
 \partial_-\phi^{\sst (s,s-1)} 
 \label{AS type action} \nn
 \eq 2\,\t(s)\,(-1)^s
 \int_{\partial M}\dd^2x\,
 \cD_\cL^{\sst (2s-1)}\phi^{\sst (s)}\,
 \partial_-\phi^{\sst (s)}\,.
\ea 
Here, $\cD_\cL^{\sst (2s-1)}$ is a differential operator of maximal order $2s-1$.
For the examples of spin-$2$ and spin-$3$, they are given by
\begin{subequations}
    \begin{align}
        & \cD_\cL^{\sst (3)} =\partial_1^{3}-2(\partial_1\,\cL +\cL\,\partial_1)\,, \\
        & \cD_\cL^{\sst (5)} =\partial_1^5-2(2\,\partial_1^3\,\cL +3\,\partial_1^2\,\cL\,\partial_1     +3\,\partial_1\,\cL\,\partial_1^2 +2\,\cL\,\partial_1^3) +8(3\,\partial_1\,\cL^2 +2\cL\,\partial_1\,\cL+3\,\cL^2\,\partial_1)\,,
    \end{align}
    \label{D35}
\end{subequations}
where $\partial_1$ and $\cL=\cL(x^+)$ both act as operators, \textit{e.g.} $\partial_1\,\cL\,\phi=\partial_1(\cL\,\phi)$,
when acting on $\phi$.
For constant $\cL$, the operator $\cD^{\sst (2s-1)}_\cL$ factorizes into the simple form,
\be 
    \cD^{\sst (2s-1)}_{\cL}
    =\partial_1\prod_{n=1}^{s-1}
    (\partial_1^2-4\,\cL\,n^2)
    =\partial_1
    (\partial_1^2-4\,\cL)\,
    (\partial_1^2-4\,\cL\,2^2)\cdots 
    (\partial_1^2-4\,\cL\,(s-1)^2)\,.
    \label{factorization}
\ee 
See Appendix \ref{sec: factorization}
for the proof of the formula.
Interestingly, these differential operators coincide with the Bol operators \cite{bol1949invarianten} (see also \cite{Gieres:1991sd,Gieres:1993my}),
which are invariant under the projective action of $SL(2,\mathbb R)$.

Recall that we should also take the gauge symmetry \eqref{gauge} into account.
For $v\propto \dd x^0$, this becomes
\be 
    \Dd_1 \phi^{\sst (s,n)}_{\rm g}=0
    \quad [1-s\le n\le s-1]\,,
\ee 
where $\phi^{\sst (s,n)}_{\rm g}$ is the gauge shift part of $\phi^{\sst (s,n)}$.
Since these conditions overlap with the asymptotic AdS conditions \eqref{const} for $n=2-s,\ldots,s-1$, the gauge degrees of freedom are concentrated in the solution of the $n=1-s$ equation
\be 
    \Dd_1\phi^{\sst (s,1-s)}_{\rm g}
    =\frac{(-1)^{s-1}}{(2(s-1))!}\,
    \cD^{\sst (2s-1)}_\cL\,\phi^{\sst (s)}_{\rm g}=0\,.
\ee 
We can see that for $2\pi$-periodic $x^1=\theta$, the kernel of $\cD^{\sst (2s-1)}_\cL$ is one-dimensional for generic (\textit{e.g.} irrational) constant $\cL$.
However, \textit{e.g.} for the cases with $\cL=-k^2/4$ for some $k\in \mathbb N$, the kernel becomes $2s-1$ dimensional.
In these cases, the origin has the conical surplus of the angle $2\,\pi\,k$. 
For the non-linear theory with $PSL(N,\mathbb R)$ symmetry, these eventually would correspond to $U(1)^{N-1}$ or 
$PSL(N,\mathbb R)^{\sst (k)}$ gauge symmetry.
Besides theses two cases, there exist other values of $\cL$ leading to non-trivial zero modes.
We discuss such cases in Section \ref{sec: mode}. 

Let us also remark that it is actually sufficient to impose the $s-1$ constraints with positive $n>0$ in \eqref{const}, as the terms involving $\phi^{\sst (s,n)}$ with $n<0$ are total derivatives. 
To see this, we first rewrite the action \eqref{FJ type} as
\begin{align}
    S_{\rm L}
    \propto &\int \dd^2 x\,
    \Big[ \partial_1 \phi^{\sst (s,0)}\partial_- \phi^{\sst (s,0)}+ s\,\cL\,(\phi^{\sst (s,1)}\partial_- \phi^{\sst (s,0)}-\phi^{\sst (s,0)}\partial_- \phi^{\sst (s,1)})\\
    &+\sum_{m=1}^{s-1} \frac{(-1)^m\,(s)_m}{(s-m)_m} \left(\partial_1(\phi^{\sst (s,-m)}\,\partial_- \phi^{\sst (s,m)})-\phi^{\sst (s,-m)}\partial_-\Dd_1\phi^{\sst (s,m)}+\partial_-\phi^{\sst (s,-m)}\Dd_1\phi^{\sst (s,m)}\right)\Big]\,,
    \nonumber
\end{align}
with the Pochhammer symbol $(a)_n=\frac{\Gamma(a+n)}{\Gamma(a)}$. 
The second line is composed of total-derivative terms and ones that vanish upon imposing only the asymptotic AdS conditions \eqref{const} with $n>0$. 
Therefore, by imposing only half of the asymptotic AdS conditions, the action reduces to the first line, which is a functional of $\phi^{\sst (s,0)}$ and $\phi^{\sst (s,1)}$, and these two fields themselves are given through $\phi^{\sst (s,s-1)}$ by the same half of the asymptotic AdS conditions.

In the end, we find that the physical degrees of freedom are in the chiral scalar $\partial_-\phi^{\sst (s)}_{\rm ph}=0$ carrying the $\cD(s,+s)$
representation of $\mathfrak{so}(2,2)$, isomorphic to the discrete series representation of $\mathfrak{sl}(2,\mathbb R)$ with the highest weight $L_0=2s$.
These representations can be interpreted as massive scalars in AdS$_2$ with definite-sign energy or tachyonic scalars in dS$_2$ with definite-sign momentum\footnote{See the recent work \cite{Farnsworth:2024yeh} for the hidden conformal symmetry of the dS$_2$ scalar.}.

For general $v$, we can consider the following generalization of the asymptotic AdS conditions,
\be
    v\wedge a^{\sst (s,n)}=\cO(z^{1-n})
    \qquad 
    [2-s\le n \le s-1]\,.
    \label{cov asymp AdS}
\ee
These conditions reduce to
\be
    v\wedge \Dd\phi^{\sst (s,n)}=0 \qquad 
    [2-s\le n \le s-1]\,,
\ee
while the gauge degrees of freedom are in the solution of
\be 
      v\wedge \Dd\phi_{\rm g}^{\sst (s,1-s)}=0\,.
\ee 
We can implement the asymptotic AdS conditions to the action \eqref{cov action} as constraints in the form of
\be
    S_{\rm L} = \int_{\partial M} \tr\left[-\frac12\, (\Dd\phi+v\,\rho)\wedge \star (\Dd\phi+v\,\rho) 
    +v \wedge(\rho+\l)\,\Dd\phi
    \right]\,,
\ee 
where the Lagrange multiplier $\l$ is given as 
\be 
    \l=\sum_{n=1-s}^{s-2}
    \l^{\sst (s,n)}\,W^{(s)}_n\,,
    \label{asym l}
\ee     
without the $W^{(s)}_{s-1}$ component.
Again, we can restrict $\lambda^{\sst (s,n)}$ to the values
$n=1-s,\ldots, -1$ since the constraints imposed by $\lambda^{\sst (s,n)}$ with $n=0,\ldots, s-2$ concern only about total derivative terms.

\section{Edge mode action of HS gravity around BTZ blackhole}
\label{sec: nonlinear}

Let us consider HS gravity formulated as $\mathfrak{sl}(N,\mathbb R)\oplus \mathfrak{sl}(N,\mathbb R)$ CS theory with boundary terms. 
The action is given by
\be 
    S_{\rm HSG}=S_{\rm L}[\cA]-S_{\rm R}[\tilde\cA]\,,
\ee 
with the left and right parts given by
\begin{subequations}
    \label{SL nonlinear}
    \begin{align}
        &S_{\rm L}[\cA] =\int_M\tr \left(\cA\wedge\dd  \cA+\frac23\,\cA\wedge \cA\wedge \cA\right)-\frac12\int_{\partial M} \tr\left(\cA\wedge\star\cA\right), \\
        &S_{\rm R}[\tilde\cA] =\int_M\tr \left(\tilde\cA\wedge\dd  \tilde\cA+\frac23\,\tilde\cA\wedge \tilde\cA\wedge \tilde\cA\right)+\frac12\int_{\partial M} \tr\left(\tilde\cA\wedge\star\tilde\cA\right)
    \end{align}
\end{subequations}
respectively.
The gauge field $\cA$ contains the gravity part $A^{\sst (n)}$ as well as
the Fronsdal fields $a^{\sst (s,n)}$ of spin $s=3,\ldots, N$\,:
\be 
    \cA=\sum_{n=-1}^1\,A^{\sst (n)}\,L_n+
    \sum_{s=3}^N \sum_{n=-(s-1)}^{s-1}\,a^{\sst (s,n)}\, W^{\sst (s)}_n\,.
\ee 
The right gauge field $\tilde\cA$ has an analogous expansion.
Together, this action describes massless fields of spin $s=2,\dots, N$ with HS self-interactions due to the non-trivial Lie brackets $[W^{\sst (s)}_n, W^{\sst (s')}_m]$. 

Note here that the boundary term still makes use of the Hodge dual defined with respect to the flat zweibein $\mathsf e^\a$, while the theory now describes a dynamical metric. 
For physical consistency, we ought to limit the boundary value of $\cA$ and $\tilde\cA$ in a way that the 3d metric has flat induced metric on the 2d boundary.

\subsection{Boundary Terms}

Let us make a few comments about the boundary term, $-\frac12\,\tr(\cA\wedge \star \cA)$, which preserves the boundary Lorentz symmetry\footnote{See \textit{e.g.} \cite{Arcioni:2002vv} for a more focused discussion about boundary terms.}.
When we rewrite the CS action in its Hamiltonian (first-order) form,
\be 
    S_{\rm Ham}=
    2\int_M\dd^3 x\,\tr\left[
    \cA_2\,\partial_0\,\cA_1
    +\cA_0\,\cF_{12}
    \right],
\ee
we obtain a boundary term from
\be
    S_{\rm CS}=
    S_{\rm Ham}+\int_{\partial M}
    \dd^2x\,\tr(\cA_0\,\cA_1)\,,
\ee 
This boundary term $\tr(\cA_0\,\cA_1)$ combines with the other boundary term $-\frac12\,\tr(\cA\wedge \star \cA)
=\frac12\,\tr[-(\cA_0)^2+(\cA_1)^2]$ to give
\be 
    S_{\rm bd}=\int_{\partial M}
    \dd^2x\,\tr\left[\cA_0\,\cA_1+
    \frac12\left(
    -(\cA_0)^2+(\cA_1)^2\right)\right],
\ee 
which is the total boundary term in Hamiltonian form: $S_{\rm L}=S_{\rm Ham}+S_{\rm bd}$. 
In the Hamiltonian treatment, the component $\cA_0$ is a non-dynamical field that can be decomposed as $\cA_0=\L+\rho$. 
To avoid redundancies within the decomposition, we require $\L$ to vanish on the boundary, while the bulk profile of $\rho$ be specifically fixed in terms of its boundary value. 
Then the variation with respect to $\cA_0$ gives
\be 
    \delta S_{\rm L}
    =2\int_{M} \dd^3x\,
    \tr\left(\delta(\L+\rho)\,\cF_{12}\right)
    +\int_{\partial M} \dd^2x\,
    \tr\left[
    \delta\rho\left(-\rho+\cA_1\right)\right].
\ee 
This has two parts: in the bulk, the $\L$ variation sets the constraint $\cF_{12}=0$; on the boundary, the $\rho$ variation sets its value as $\rho=\cA_1$.
Note here that the variation $\delta \rho$ in the bulk is not an arbitrary bulk field as its bulk profile is simply a fixed function.
In the end, we find that the bulk variation $\delta \L$ and the boundary variation $\delta\rho$ are independent.
If we perform first the boundary variation $\delta\rho$, the boundary term will be reduced to
\be 
    S_{\rm bd}=\int \dd^2\,\tr
    \left[(\cA_1)^2\right],
\ee     
\textit{i.e.} the typical (additional) boundary term of the Hamiltonian treatment of CS theory.
On the other hand, performing the bulk variation first corresponds to the reduction procedure \cite{Arvanitakis:2022bnr} that we follow in this paper.
The boundary variation can be performed eventually after such a boundary reduction, and this step is nothing but 
the conversion of the covariant form \eqref{WZWM} of the action into the FJ form \eqref{FJ type}. 

\subsection{Equations of motion}

The non-linear equation of motion is also just the flatness condition $\dd \cA+\cA\wedge \cA=0$ and can be solved by 
\be 
    \cA = g_\varphi^{-1}\,\underline{\cA}\,g_\varphi+
    g_\varphi^{-1}\,\dd g_\varphi\,.
\ee
Note here that we singled out the BTZ background 
\ba 
    \underline{\cA}
    \eq -\frac{\dd z}z\,L_0+\left(\frac1z\,L_{+1}-z\,\cL\,L_{-1}\right)\dd x^+ \nn
    \eq 
    e^{-\rho\,L_0}\,
    e^{-(L_{+1}-\cL\,L_{-1})\,x^+}
    \dd \left[e^{(L_{+1}-\cL\,L_{-1})\,x^+}
     e^{\rho\,L_0}\right],
\ea
where $z=e^{-\rho}$ and $\cL$ is from now on assumed to be a constant (the non-constant part of $\cL$ can be always absorbed into $g_\varphi$).
The on-shell gauge field can be expressed in pure gauge form as 
\be 
    \cA=G^{-1}\,\dd G\,,
\ee 
with an $SL(N,\mathbb R)$ element 
\be 
    G=\underline{g}\,e^{\rho\,L_0}\,g_\varphi=\underline{g}\,
    g_\phi\,e^{\rho\,L_0}\,,
\ee 
where 
\be 
    \underline{g}=e^{(L_+-\cL\,L_-)\,x^+},\qquad g_\varphi=e^{-\rho\,L_0}\,g_\phi\,e^{\rho\,L_0}
\ee 
are the gauge function associated with the background field $\underline{\cA}$ up to the $z$-dependence and the non-linear counterpart of the Fronsdal fields (including graviton) $\varphi$ respectively.
In contrast, $g_\phi$ is a $z$-independent field and the non-linear counterpart of the boundary value $\phi$.
From now on, we use
\be 
    g=\underline{g}\,g_\phi\,,
\ee 
and hence $G=g\,e^{\rho\,L_0}$\,.
The boundary equation of motion $(\cA-\star \cA)|_{\partial M}=G^{-1}\partial_- G\,\big|_{\partial M}=0$ simply becomes 
\be 
    g^{-1}\,\partial_- g=0\,,
\ee 
so on-shell we have chiral field $g=g(x^+)$.
The asymptotic AdS condition \cite{Campoleoni:2010zq,Bunster:2014mua,Campoleoni:2014tfa} is 
\be 
    \cA^{\sst (s,n)}_1\,|_{\partial M}=\underline{\cA}^{\sst (s,n)}_{1}\,|_{\partial M}
    \qquad [n>1-s]\,,
\ee 
which can be recast into a condition on $g$: 
\be 
    \left[g^{-1}\, \partial_1g
    =L_{+1}\right]^{\sst (s,n)}
    \qquad [n>1-s]\,.
    \label{nl aads}
\ee 
Since we have separated the background part $\underline{g}$, the fluctuation part $g_\phi$ is connected to the identity. Hence, the above condition can be solved by parametrizing $g_\phi$ using any decomposition of $SL(2,\mathbb R)$ which provides a good coordinate chart near the identity of $SL(2,\mathbb R)$.

\subsection{Boundary reduction}

Let us proceed with the covariant reduction of the bulk action by using once again the decomposition, 
\be
    A=B+v\,\psi\,.
\ee 
As before, $\psi$ acts as a Lagrange multiplier and enforces
\begin{equation}
    v \wedge (\dd B+B\wedge B) = 0\,.
\end{equation}
This is the flatness condition on the full gauge field (background and fluctuations) up to terms proportional to $v$, so the solution is (as shown in \cite{Arvanitakis:2022bnr})
\begin{equation}
    A = G^{-1}\dd G + v\,\rho\,,
\end{equation}
where  $\rho$ takes value
in $\mathfrak{sl}(N,\mathbb R)$.
Plugging this into the action \eqref{SL nonlinear}, we find
the democratic action,
\be 
    S_{\rm L}= \int_{\partial M} 
    \tr\left[
    -\frac12\,(g^{-1}\dd g+v\,\rho)
    \wedge \star (g^{-1}\dd g+v\,\rho)
    +v\wedge (\rho+\l)\, g^{-1}\dd g\right]
    +S_{\rm WZW}[g]\,,
    \label{cov act}
\ee
where the $z$-dependence decoupled and the asymptotic conditions, 
\be 
    v\wedge \cA^{\sst (s,n)}\,|_{\partial M}=
    v\wedge \underline{\cA}^{\sst (s,n)}
    \qquad [n>1-s]\,,
\ee 
are implemented as constraints with Lagrange multiplier $\l\in\mathfrak{sl}(N,\mathbb R)$ satisfying \eqref{asym l}.
Note again that the conditions can be relaxed to the nilpotent algebra with $n>0$, because the others are only related to total derivative terms.
The last term is the usual WZW term given by
\be 
    S_{\rm WZW}[g]
    =-\frac13\int_{M}
    \tr\left[(g^{-1}\dd g)^3\right].
\ee 
For $v\propto \dd x^0$, this action becomes a gauged WZW model.
It is also interesting that this form of the action does not have an explicit dependence on the background parameter $\cL$\,.
The dependence arises purely from our choice of the parameterization for $g=\underline{g}\,g_\phi$.

\subsection{Non-covariant expression}

In the previous section, we identified the chiral action \eqref{cov act} obtained by reducing the half of $SL(N,\mathbb R)$ CS (HS) gravity. 
This action has manifest Lorentz covariance thanks to the one-form field $v$, which is pure gauge.
In this section, we show how by gauge fixing $v$, the action \eqref{cov act} further reduces to the \emph{geometric action}, namely the AS action and its HS generalization when $N=2$ or $N=3$ respectively.
Compared to the early works \cite{Bershadsky:1989mf,Marshakov:1989ca,Bilal:1990wn}, we particularly emphasize the role of the background solution.

By choosing the gauge $v\propto \dd x^0$ and integrating out $\rho$ and $\l$, the covariant action \eqref{cov act} reduces to the chiral WZW action (the non-linear counterpart of the FJ-type action \cite{Sonnenschein:1988ug}),
\be 
    S_{\rm L}=-2\int_{\partial M} \dd^2x\,\tr\left(g^{-1}\partial_1 g\,
    g^{-1}\partial_- g\right)
    -\frac13\int_{M} \tr\left[(g^{-1}\dd g)^3\right],
    \label{chWZW}
\ee 
supplemented with the asymptotic AdS condition \eqref{nl aads}.
The asymptotic AdS condition allows us to express $\phi^{\sst (s,n)}$ with $n<s-1$ in terms of $\phi^{\sst (s,s-1)}$.

\subsubsection*{Gravity}

Let us briefly review the gravity case, where solving the asymptotic AdS condition leads to the AS action.

As before, we shall consider the field $\cA$ given by the gauge function $g_\phi$ connected to the identity on top of the background $\underline{\cA}$. 
Then, consider the decomposition
\be
    g_\phi=e^{\varepsilon\,(L_{+1}-\cL\,L_{-1})}
 \,e^{\s\,L_0}\,
 e^{f\,L_{-1}}\,,
     \label{iwazawa decomp}
\ee 
whose linearization gives the gravitational fluctuation as
\be 
    \varepsilon=\phi^{\sst (1)}+\cO(\phi^2)\,,
    \quad 
    \s=\phi^{\sst (0)}+\cO(\phi^2)\,,
    \quad 
    f=\phi^{\sst (-1)}+\cL\,\phi^{\sst (1)}
    +\cO(\phi^2)\,.
\ee 
The advantage of the decomposition \eqref{iwazawa decomp} is that it leads to a simple expression for $g$:
\be 
    g=e^{\chi\,(L_{+1}-\cL\,L_{-1})}
 \,e^{\s\,L_0}\,
 e^{f\,L_{-1}}\,.
\ee 
We see that the field $\varepsilon$ combines nicely with the background $x^+$ in $\chi=x^++\varepsilon$.
For further explicit calculation, we use the fundamental representation of $\mathfrak{sl}(2,\mathbb R)$,
\be
    L_0=\frac12\begin{pmatrix} 1 & 0 \\ 0 & -1 \end{pmatrix}\,, \qquad L_{+1}=\begin{pmatrix} 0 & 0 \\ 1 & 0 \end{pmatrix}\,,
    \qquad
    L_{-1}=\begin{pmatrix} 0 & -1 \\ 0 & 0 \end{pmatrix}\,,
\ee 
with which the gauge function $g$ takes the form,
\be 
g= 
\begin{pmatrix} \cosh(\cL^{\frac12}\,\chi) & \cL^{\frac12} \sinh(\cL^{\frac12}\,\chi) \\ \cL^{-\frac12}\sinh(\cL^{\frac12}\,\chi) & \cosh(\cL^{\frac12}\,\chi) \end{pmatrix}
		\begin{pmatrix} e^{\frac12\,\s} & 0 \\ 0 & e^{-\frac12\,\s} \end{pmatrix}
		\begin{pmatrix} 1 & -f \\ 0 & 1 \end{pmatrix}.
  \label{iwzw}
\ee
From this expression, we can verify that for $\cL<0$ (``space-like'' $\chi$), the gauge function $g$ becomes periodic.

For the AdS background with 
$\cL=\tilde\cL=-\frac14$, 
the gauge functions become periodic
up to $\mathbb Z_2$ equivalence, $g(x^++2\pi)=-g(x^+)$
and $\tilde g(x^-+2\pi)=-\tilde g(x^-)$, and this leads to
trivial holonomy.
For different values of $\cL <0$
with $\cL\,\tilde\cL>0$, the holonomy is non-trivial, reflecting that the background geometry has a conical singularity at the origin.
For the BTZ black hole case with $\cL,\tilde\cL>0$ (``time-like'' $\chi$), 
by moving to the Euclidean
setting with  $t_{\sst\rm E}=i\,t$ and $r_{\sst\rm E}=-i\,r_-$
where
the background
$SL(2,\mathbb C)$ connection $\underline{A}$ is determined by $\cL=(r_++i\,r_{\sst\rm E})^2$,
the holomorphic gauge function $g$ has the periodicity,
\be 
 g(\theta+2\pi\,{\rm Re}(\t),t_{\rm\sst E}+2\pi\,{\rm Im}(\t))=-g(\theta,t_{\rm\sst E})\,,
 \label{periodicity}
\ee 
where $\t$ is given by
\be
    \t=\frac{2\pi\, i}{\cL^{\frac12}}
    =\frac{2\pi\,i}{r_++i\,r_{\rm\sst E}}
    =\frac{2\pi(-r_{\rm\sst E}+i\,r_+)}{r_+{}^2+r_{\rm\sst E}{}^2}\,.
\ee 
From this, we find $\b=|{\rm Im}(\t)|=\frac{2\pi\,r_+}{r_+{}^2+r_{\rm\sst E}{}^2}
=\frac{2\pi\,r_+}{r_+{}^2-r_{-}{}^2}$.

Let us now return to the boundary reduction.
From the decomposition \eqref{iwzw}, we get 
\be
	g^{-1}\dd g =
    \a\,L_1+\d\,L_0+\g\,L_{-1}=
    \begin{pmatrix}
        \frac \d2 & -\g\\
        \a & -\frac \d2
    \end{pmatrix}\,,
\ee 
with 
\ba
&\a =  e^{\s}\,\dd \chi\,,
\qquad 
\d=\dd \s+2\,e^{\s}\,f\,\dd \chi\,,
    \nn 
&	\g= \dd f+f\,\dd\s +
  e^{\s}\,f^2\,\dd \chi
  -\cL\,e^{-\s}\dd \chi\,, 
  \label{sl2}
\ea
and the asymptotic AdS conditions,
$\a_1=1$ and $\d_1=0$, reduce to 
\be 
    e^{-\s}=\partial_1\chi\,,
    \qquad 
    f=-\frac12\,\partial_1\s\,.
    \label{asymp cond grav}
\ee 
In this decomposition, the chiral WZW action \eqref{iwzw} reads
\be 
     S_{\rm L}=
     \int_{\partial M} \dd^2x
     \left[-\,\partial_1\s\,\partial_-\s+2\,e^\s(\partial_1\chi\,\partial_-f
     +\partial_-\chi\,\partial_1f)
     -4\,\cL\,\partial_1\chi\,\partial_-\chi\right]
     +e^\s \dd \chi\wedge \dd f\,.
\ee 
Upon imposing the first of the asymptotic AdS conditions \eqref{asymp cond grav}, the action reduces to the AS action \cite{Alekseev:1988tj,Alekseev:1988vx,Alekseev:1988ce},
\begin{equation}
        S_{\rm L} 
        =\int_{\partial M}\dd^2x
        \left(
        -\frac{\partial_1^2\chi\,\partial_-\partial_1\chi}{(\partial_1\chi)^2}-4\,\cL\,\partial_1\chi\,\partial_-\chi\right)
\end{equation}
as expected.

\subsubsection*{$SL(3,\mathbb R)$ HS gravity}

Let us consider now the case of $SL(3,\mathbb R)$ HS gravity.
Again, the problem reduces to finding a suitable parameterization for the gauge function $g_\phi\in PSL(3,\mathbb R)$ to impose the asymptotic AdS condition.
As we have emphasized before, after separating the background $\underline{\cA}$, the gauge function $g_\phi$ is connected to the identity, and hence we may choose any decomposition that defines a good coordinate system near the identity. 
Here, we choose $g= g_2\,g_3$ with 
\begin{subequations}
    \label{decomp g3}
    \begin{align}
        &g_2=e^{\chi\,(L_{+1}-\cL\,L_{-1})} \,e^{\s\,L_0}\, e^{f\,L_{-1}}\,, \\
        &g_3= e^{\phi^{\sst (3,2)}\,W^{(3)}_2+\phi^{\sst (3,1)}\,W^{(3)}_1}\,
e^{\phi^{\sst (3,0)}\,W^{(3)}_0}\,
 e^{\phi^{\sst (3,-1)}\,W^{(3)}_{-1}+\phi^{\sst (3,-2)}\,W^{(3)}_{-2}}\,.
    \end{align}  
\end{subequations}
In the fundamental representation of $\mathfrak{sl}(3,\mathbb R)$, we take
\be 
    L_{+1}=\begin{pmatrix}
    0 & 0 & 0\\
    \sqrt{2} & 0 & 0\\
    0 & \sqrt{2} & 0
	\end{pmatrix}, \qquad 
  L_{0}=\begin{pmatrix}
    1 & 0 & 0 \\
    0 & 0 & 0 \\ 
    0 & 0 & -1
	\end{pmatrix}, \qquad
  L_{-1}=\begin{pmatrix}
    0 & -\sqrt{2} & 0 \\ 
    0 & 0 & -\sqrt{2} \\
    0 & 0 & 0
	\end{pmatrix}, 
\ee 
and
\begin{gather}
    W^{(3)}_{+2}=\begin{pmatrix}
    0 & 0 & 0\\
    0 & 0 & 0\\
    2 & 0 & 0
	\end{pmatrix}, \qquad 
  W^{(3)}_{0}=\begin{pmatrix}
    \frac13 & 0 & 0 \\
    0 & -\frac23 & 0 \\ 
    0 & 0 & \frac13
	\end{pmatrix}, \qquad
  W^{(3)}_{-2}=\begin{pmatrix}
    0 & 0 & 
    2 \\ 
    0 & 0 & 0 \\
    0 & 0 & 0
	\end{pmatrix}, \\
       W^{(3)}_{+1}=\begin{pmatrix}
    0 & 0 & 0\\
    \frac1{\sqrt{2}} & 0 & 0\\
    0 & -\frac1{\sqrt{2}} & 0
	\end{pmatrix}, \qquad 
  W^{(3)}_{-1}=\begin{pmatrix}
    0 & \frac1{\sqrt{2}} & 0 \\
    0 & 0 & -\frac1{\sqrt{2}} \\ 
    0 & 0 & 0
	\end{pmatrix}.
\end{gather}    
Then the asymptotic AdS condition,
\be 
    g^{-1}\partial_1 g
    =g_3^{-1}\,g_2^{-1}\partial_1 g_2\,g_3
    +g_3^{-1}\,\partial_1 g_3
    =
    \begin{pmatrix}
   0 & -\sqrt{2}\,l & 2\,w \\
   \sqrt{2} & 0 & -\sqrt{2}\,l \\
    0 & \sqrt{2} & 0 
    \end{pmatrix}\,,
\ee
boils down to the following six equations:
\be
    \alpha_1
    =\cosh(\phi_0)- \frac34\,\gamma_1\,  \phi _1^2\,,
    \qquad \d_1=0\,,
\ee
\be 
    \phi_1=-
    \frac{
    \partial_1\phi_2
     +\frac14\,\g_1\,\phi_1^3}{\a_1}
\,,\qquad
\sinh(\phi _0)=-\frac{\partial_1 \phi_1-4 \,\gamma_1  \,\phi _2}2\,,
\ee
\be 
\phi_{-1}=
   -\frac{\partial_1\phi_0
   -3\,\gamma_1\,\phi _1}{3}\,,
   \qquad 
  \phi _{-2}=-\frac{\partial_1\phi_{-1}-
  2\,\g_1\,\sinh(\phi _0)}4\,,
\ee 
together with
\be
    -l=-\g_1\,\cosh(\phi_0)
    +\frac34\,\phi_{-1}^2\,,
    \qquad 
    w=\partial_1\phi_{-2}
    -\g_1\,\cosh(\phi_0)\,\phi_{-1}
    +\frac12\,\phi_{-1}^3\,.
\ee
Here, $\a, \d, \g$ are given in \eqref{sl2}, and the fields are relabeled as $\phi_n=\phi^{\sst (3,n)}$ for simplicity.

These six equations are the non-linear counterpart of \eqref{recur} with $\g_1=-\cL$, so the fields $f$, $\sigma$, and $\phi_{n}$ for $n=-2,\dots 1$ can be perturbatively determined  in terms of $\chi$ and $\phi_{2}$, as expected.
The edge mode action is in principle determined by implementing these relations in the chiral WZW action 
\eqref{chWZW}.

\section{Edge mode action of HS gravity with HS charge}
\label{sec: HS charge}

In HS gravity, we can also consider a solution with a non-trivial HS charge.
In the case of vanishing chemical potential,\footnote{When a background solution has a nonzero chemical potential, it modifies the boundary condition as well as the boundary term. 
Therefore, it does not fit in with the present set-up. Note that the HS black holes considered in \cite{Gutperle:2011kf, Bunster:2014mua} are the typical solutions with non-trivial chemical potential.} such a solution\footnote{A solution with non-trivial HS charge but zero chemical potential might be interpreted as a kind of 
generalized conical defect.
See e.g. \cite{Castro:2011iw,Campoleoni:2017xyl} for detailed analysis and discussions of the conical defect type solutions in HS gravity.
} also satisfies the asymptotic AdS condition.  
For $PSL(3,\mathbb R)$, it is given by
\be 
    \underline{\cA}=
    e^{-\rho\,L_0}\left(L_{+1}-\cL\,L_{-1}+\cW\,W^{(3)}_{-2}\right)\dd x^+\,
    e^{\rho\,L_0}\,
\ee 
with $\cL=\cL(x^+)$ and $\cW=\cW(x^+)$.
Since the covariant form of the corresponding edge mode action does not have an explicit dependence on $\underline{\cA}$, the action \eqref{cov act} is still valid, while the gauge function should be parameterized in a way that it is connected to $\underline{g}$. 
Making use of an appropriate gauge function, we can obtain an explicit expression for the edge mode action in a non-covariant form, but such an expression would not be very illuminating --- in either case, the expression depends on the choice of the parametrization. 
We shall therefore focus on linear fluctuations around a background with a non-trivial HS charge.

\subsection{Linearized edge mode action around HS background}

The fluctuation fields $\phi^{\sst (s,n)}$ are subject to the asymptotic AdS conditions which for $v\propto \dd x^0$\,  become
\ba 
   && \Dd_1\phi^{\sst (2,1)}=\partial_1 \phi^{\sst (2,1)}+ \phi^{\sst (2,0)}=0\,,\nn
   && \Dd_1\phi^{\sst (2,0)}=\partial_1 \phi^{\sst (2,0)}+ 2\,\phi^{\sst (2,-1)}+2\,\cL\,\phi^{\sst (2,1)}-4\,\cW\,\phi^{\sst (3,2)}=0\,,
\ea 
and 
\ba
   && \Dd_1\phi^{\sst (3,2)}=\partial_1 \phi^{\sst (3,2)}+
   \phi^{\sst (3,1)}=0\,,\nn
   && \Dd_1\phi^{\sst (3,1)}=\partial_1 \phi^{\sst (3,1)}+ 2\,\phi^{\sst (3,0)}+4\,\cL\,\phi^{\sst (3,2)}=0\,,\nn
   && \Dd_1\phi^{\sst (3,0)}=\partial_1 \phi^{\sst (3,0)}+ 
   3\,\phi^{\sst (3,-1)}
   +3\,\cL\,\phi^{\sst (3,1)}
   =0\,,\nn
   && \Dd_1\phi^{\sst (3,-1)}=\partial_1 \phi^{\sst (3,-1)}+ 4\,\phi^{\sst (3,-2)}
   +2\,\cL\,\phi^{\sst (3,0)}
   +4\,\cW\,\phi^{\sst (2,1)}=0\,,
\ea 
for spin-2 and spin-3 respectively.
Solving these equations, we find
\ba
       \Dd_1\phi^{\sst (2,-1)}\eq \partial_1 \phi^{\sst (2,-1)}+ 3\,\phi^{\sst (2,0)}+\cL\,\phi^{\sst (2,1)}-\cW\,\phi^{\sst (3,1)}\nn
       \eq \frac12\,\cD^{\sst (3)}_\cL\,\phi^{\sst (2)}
       +(\partial_1\,\cW
       +\cW\,\partial_1)\,\phi^{\sst (3)}\,\nn
       \Dd_1\phi^{\sst (3,-2)}\eq\partial_1 \phi^{\sst (3,-2)}+ 5\,\phi^{\sst (3,-1)}
   +\cL\,\phi^{\sst (3,0)}
   +2\,\cW\,\phi^{\sst (2,0)} \nn
   \eq \frac1{24}\,\cD^{\sst (5)}_\cL\,\phi^{\sst (3)}
   -2(\partial_1\,\cW+\cW\,\partial_1)\,\phi^{\sst (2)}\,,
\ea 
so the FJ-type action \eqref{FJ type} reduces to
\ba  
    S_{\rm L}
    \eq \int_{\partial M} \dd^2x\left(
    2\,\Dd_1\phi^{\sst (2,-1)}\,\partial_-\phi^{\sst (2,1)}
    -2\,
     \Dd_1\phi^{\sst (3,-2)}\,\partial_-\phi^{\sst (3,2)}\right) \nn
     \eq \int_{\partial M} \dd^2 x
    \left[
    \cD^{\sst (3)}_\cL\,\phi^{\sst (2)}\, \partial_-\phi^{\sst (2)}
    -
    \frac1{12}\,\cD^{\sst (5)}_\cL\,\phi^{\sst (3)}\,\partial_-\phi^{\sst (3)}
    -2\left[(\partial_1\,\cW+\cW\,\partial_1)\phi^{\sst (2)}\right]\partial_- \phi^{\sst (3)}\right].
    \quad 
\ea
The fields $\phi^{\sst (2)}$ and $\phi^{\sst (3)}$ satisfy higher-derivative equations of motion, which can be put into matrix form as
\be 
    \begin{pmatrix}
    \cD^{\sst (3)}_\cL
    & -(\partial_1\,\cW+\cW\,\partial_1)\\
    -(\partial_1\,\cW+\cW\,\partial_1)
    & -\frac1{12}\,\cD^{\sst (5)}_\cL
    \end{pmatrix}
    \partial_-
    \begin{pmatrix}
    \phi^{\sst (2)} \\
    \phi^{\sst (3)}
    \end{pmatrix}=0\,.
\ee
Note that these equations bear strong resemblance
to the Ward identities derived in \cite{Li:2015osa} --- \textit{cf.} equations (3.69) and (3.70) therein.
Once again, up to the gauge symmetry corresponding to the kernel of the first factor, the solutions are simply determined by the chirality conditions $\partial_-\phi^{\sst (2)}=0=\partial_-\phi^{\sst (3)}$.
It is noteworthy that, despite the simple decoupled equations, the action for $\phi^{\sst (2)}$ and $\phi^{\sst (3)}$ is not in a diagonal form, due to coupled gauge degrees of freedom. 

\subsection{Mode expansions}
\label{sec: mode}

Let us now consider the Euclidean case, where $(x^1=\theta, x^0=-i\,t_{\rm E})$ has the periodicity
\be 
   \phi^{\sst (s)}(\theta+2\pi,t_{\rm E})
   =\phi^{\sst (s)}(\theta,t_{\rm E})
   =\phi^{\sst (s)}(\theta+2\pi\,{\rm Re}(\t),
   t_{\rm\sst E}+
   2\pi\,{\rm Im}(\t))\,.
\ee 
We may therefore consider the mode expansion
\be 
    \phi^{\sst (s)}(\theta,t_{\rm E})
    =\sum_{(m,n)\neq\, {\rm zero\, modes}}
    \phi^{\sst (s)}_{m,n}\,
    e^{i\left[m
    \left(\theta-\frac{
    {\rm Re}(\t)}{
    {\rm Im}(\t)}\,
    t_{\rm\sst E}\right)
    +\frac{n}{{\rm Im}(\t)}\,t_{\rm\sst E}\right]}
\ee 
with complex coefficients $(\phi_{m,n}^{\sst (s)})^*=\phi_{-m,-n}^{\sst (s)}$.
For constant $\cL$ and $\cW$, the action becomes
\be 
    S_{\rm L}\propto
    \sum_{(m,n)\neq\, {\rm zero\, modes}}
    \begin{pmatrix} \phi_{m,n}^{\sst (2)} \\ \phi_{m,n}^{\sst (3)}
    \end{pmatrix}^\dagger
   K_{m,n}\,
      \begin{pmatrix} \phi_{m,n}^{\sst (2)} \\ \phi_{m,n}^{\sst (3)}
    \end{pmatrix}\,,
\ee 
with the kinetic operator now taking the form
\be
    K_{m,n}= \begin{pmatrix}
   m(m^2+4\,\cL)
    & 2\,m\,\cW\,\\
    2\,m\,\cW
    & \frac1{12}\,
    m(m^2+4\,\cL)(m^2+16\,\cL)
    \end{pmatrix}
    (n-m\,\bar\t)\,.
\ee 
The zero modes arise when
\be
   \det K_{m,n}=
   m^2 \left[\frac{(m^2+4\,\cL)^2(m^2+16\,\cL)}{12}-4\,\cW^2\right](n-m\,\bar\t)^2=0\,.
\ee
From this we see that $\phi^{\sst (2)}_{0,n}$ and 
$\phi^{\sst (3)}_{0,n}$, corresponding to the $U(1)\times U(1)$ symmetry, are always zero modes.
The other zero modes depend on the values of $\cL$ and $\cW$.
For $\cW=0$, the modes $\phi^{\sst (2)}_{\pm k,n}$, $\phi^{\sst (3)}_{\pm k,n}$, and $\phi^{\sst (3)}_{\pm 2k,n}$ become zero modes for $\cL=-k^2/4$ --- together with $\phi^{\sst (2)}_{0,n}$ and $\phi^{\sst (3)}_{0,n}$ they form the $PSL(3,\mathbb R)^{\sst (k)}$ symmetry.
For $\cL=-(2j+1)^2/16$, only $\phi^{\sst (3)}_{\pm (2j+1),n}$ are additional zero modes --- these might form a $U(1)\times PSL(2,\mathbb R)^{\sst (j)}$ symmetry.
Note that the non-zero modes have positive kinetic term only for $\cL\ge  -\frac1{16}$: this bound is tightened from $-\frac14$ due to the spin-3 modes, and implies that the AdS background is unstable. 
For $PSL(N,\mathbb R)$ HS gravity, the bound will be further strenghtened to $\cL\ge  -\frac1{4(N-1)^2}$, and eventually to $\cL\ge 0$ for the case of HS algebra $hs(\lambda)$ \cite{Vasiliev:1989qh,Vasiliev:1989re}: see also \cite{Gaberdiel:2011wb,Campoleoni:2011hg} for the asymptotic analysis of the $hs(\l)$ HS gravity.

For $\cW\neq 0$, the two eigenvalues of $K_{1,n}/(n-\bar \t)$,
\be 
    \frac{
    (1+4\cL)(13+16\cL)
    \pm \sqrt{(1+4\cL)^2(-11+16\cL)^2+2304\,\cW^2}
    }{24}\,,
\ee 
are non-negative for
\be 
    (1+4\,\cL)^2(1+16\,\cL)\ge 48\,\cW^2\,.
\ee 
When this equality holds, new zero modes appear.
This case corresponds to the $l=0$ point of the new class of backgrounds where $\cW$ is fixed by $\cL$ and an integer $l$ as
\be 
    \cW=\pm\sqrt{\frac{(l^2+4\cL)^2(l^2+16\cL)}{48}}
    \qquad [l \in \mathbb Z_{>0}]\,.
\ee 
These backgrounds have the zero modes,
\be 
 \begin{pmatrix} \phi_{\pm l,n}^{\sst (2)} \\ \phi_{\pm l,n}^{\sst (3)}
    \end{pmatrix}=
\begin{pmatrix}\mp\sqrt{\frac{l^2+16\cL}{12}}\\1\end{pmatrix}\,
\phi_{l,n}^{\sst (3)}\,,
\ee 
and may enjoy another class of $U(1)\times PSL(2,\mathbb R)^{\sst [l]}$ symmetry.

For the $PSL(N,\mathbb R)$ theory, we expect to see a larger variety of zero mode structures. 
They could be well captured by the analysis of coadjoint orbits of $\cW_N$ algebras: see, \textit{e.g.}, \cite{Avan:1991pi,Dhar:1992hr,Tjin:1993dk}
for the discussions on coadjoint orbits of W algebras.

\acknowledgments{
The authors are grateful to Stefan Fredenhagen and Kurt Hinterbichler for related discussions, and Stefan Theisen and Arkady Tseytlin for feedback on the draft.

The work of E. J. was supported by the National Research Foundation of Korea (NRF) grant funded by the Korean government (MSIT) (No. 2022R1F1A1074977).
C. Y.-R. C. was supported by the Imperial College London President's PhD Scholarship and National Taiwan University 113L400-1NTU OED Grant.
K.M. was supported by UKRI and STFC Consolidated Grants ST/T000791/1 and ST/X000575/1.
J. Y. was supported by the National Research Foundation of Korea (NRF) grant funded by the Korean government (MSIT) (No. 2022R1A2C1003182). 
J. Y. is supported by an appointment to the JRG Program at the APCTP through the Science and Technology Promotion Fund and Lottery Fund of the Korean Government. 
This is also supported by the Korean Local Governments - Gyeongsangbuk-do Province and Pohang City. 
J. Y. is also supported by Korea Institute for Advanced Study (KIAS) grant funded by the Korean government. 
C. Y.-R. C. and K.M. were also supported by the Higher Education and Science Committee of the Republic of Armenia, under the Remote Laboratory Program, grant number 24RL-1C047.
}

\appendix

\section{Factorization of the operator}
\label{sec: factorization}

In this section, we provide
the proof for the factorization 
formula \eqref{factorization}.
For that, we perform a $SL(2,\mathbb R)$ transformation to change 
the basis $W_n^{\sst (s)}$:
the gauge connection $a$
and the gauge function $\phi$
can be expanded as 
\be
    a=\sum_{n=1-s}^{s-1}
    a^{\sst(s,n)}\,W^{(s)}_n\,,
    \qquad 
    \phi=\sum_{n=1-s}^{s-1}
    \phi^{\sst(s,n)}\,W^{(s)}_n\,,
\ee 
and they are related by
$a=\Dd\phi$
with the differential $\Dd=\dd +\underline{A}$.
A $SL(2,\mathbb R)$ transformation $h$ 
leaves the space spanned by $W_n^{(s)}$'s invariant: 
\be 
    \tilde \phi=h^{-1}\,\phi\,h\,,
    \qquad 
    \tilde a=h^{-1}\,a\,h
    =\tilde \Dd\tilde\phi\,,
\ee 
where the transformed differential $\tilde \Dd=\dd+\underline{\tilde{A}}$ is given by $\underline{\tilde{A}}
    =h^{-1}\dd h+ h^{-1}\,\underline{A}\,h$\,.
If we consider a transformation
of the form,
\be
    h=e^{\eta\,L_{-1}}\,,
\ee 
the generator $W^{(s)}_{s-1}$ is
invariant, in particular,
the corresponding component of gauge functions are the same:
\be 
    \phi^{\sst (s,s-1)}=\tilde \phi^{\sst (s,s-1)}\,.
    \label{phi iden}
\ee
Moreover, the asymptotic AdS condition
is also invariant:
\be 
    a_1^{\sst (s,n)}=0\quad [2-s\le n \le s-1]
    \quad \Longleftrightarrow \quad 
    \tilde a_1^{\sst (s,n)}=0\quad [1-s\le n \le s-2]\,,
\ee 
and as a result, the remaining
components are the same:
\be 
    a_1^{\sst (s,1-s)}=
    \tilde a_1^{\sst (s,1-s)}\,.
    \label{a iden}
\ee

In our case, the background gauge connection is
$\underline{A}=(L_1-\cL\,L_{-1})\,\dd x^+$, and the transformed background gauge connection becomes
\be
    \underline{\tilde{A}}=
     \left[L_1+2\,\eta\,L_0+(\eta'+\eta^2-\cL)\,L_{-1}\right]\dd x^+\,.
\ee
By imposing the condition,
\be 
    \eta'+\eta^2=\cL\,,
    \label{eta eq}
\ee
the  non-vanishing components
of $\underline{\tilde{A}}$
are adjacent,
and the asymptotic AdS condition
becomes
\ba
    && \left(\partial_1 - 2\,n\,\eta \right)\tilde\phi^{\sst (s,n)} + (s-n)\,\tilde\phi^{\sst (s,n-1)}=0 \qquad 
    [2-s\le n\le s-1]\,,\\
   && \left[\partial_1 - 2\,(1-s)\,\eta \right] \tilde\phi^{\sst (s,1-s)} =\tilde a_1^{\sst (s,1-s)}\,.
\ea
The first equation allows
to express 
$\tilde\phi^{\sst (s,n)}$ in term of $\tilde \phi^{\sst (s,n+1)}$ as
\be
    \tilde\phi^{\sst (s,n)}
    = 
    \frac{(\partial_1 - 2\,(n+1)\,\eta )\,\tilde\phi^{\sst (s,n+1)}}{s-1-n}
    \qquad [1-s\le n\le s-2]\,,
\ee
and, the gauge connection $\tilde a_1^{\sst (s,1-s)}$ can be found to be
\ba
    \tilde a_1^{\sst (s,s-1)}\eq   {1 \over (2s-2)!} {\displaystyle \prod_{j=1}^{2s-1}}\big[\partial_1 - 2(-s+j)\,\eta \big]\,\tilde\phi^{\sst (s,s-1)}\nn
    \eq  {1\over (2s-2)!}\,\partial_1\prod_{n=1}^{s-1} (\partial_1^2 -4\,n^2\,\eta^2)\,\tilde\phi^{\sst (s,s-1)}\,.
\ea
Thanks to \eqref{phi iden} and \eqref{a iden}, the final result
is also valid for 
$a_1^{\sst (s,s-1)}$ and 
$\phi^{\sst (s,s-1)}$.
For a constant $\cL$,
$\eta^2=\cL$, so we
obtain the formula \eqref{factorization}.

\bibliography{biblio}
\bibliographystyle{JHEP}

\end{document}